\documentclass[showpacs,preprintnumbers,nofootinbib,amsmath,amssymb]{revtex4}
\usepackage{hyperref,slashed,color}
\usepackage{graphicx,subfig}
\usepackage{ulem}
\allowdisplaybreaks

\begin{document}
\preprint{TUHEP-TH-13178}
\title{Amending the Vafa-Witten Theorem}

\bigskip
\author{Chuan Li$^{2}$\footnote{Email:\href{mailto:lcsyhshy2008@yahoo.com.cn}{lcsyhshy2008@yahoo.com.cn}.}
 and Qing Wang$^{1,2}$\footnote{Email:
\href{mailto:wangq@mail.tsinghua.edu.cn}{wangq@mail.tsinghua.edu.cn}.}\footnote{corresponding
author}\\~}

\bigskip
\affiliation{$^1$Center for High Energy Physics, Tsinghua University, Beijing 100084, People's Republic of China\\
$^2$Department of Physics, Tsinghua University, Beijing 100084,
People's Republic of China\footnote{mailing address}}

 \begin{abstract}
 The strong version of the Vafa-Witten theorem is shown may not to hold because the zero condensate from a direct computation of the order parameter is found to be a result on the symmetric vacuum. The validity of the Vafa-Witten theorem relies then on its weak version, that the Goldstone boson is absent in vector-like gauge theories with vanishing $\theta$ angle. The existence of a charged $\rho$ meson condensate, which violates electromagnetic gauge symmetry, is consistent with this weak version of the Vafa-Witten theorem when applied to strong magnetic fields in QCD.
 \end{abstract}
 \maketitle

The behavior of the vacuum is a very important problem in quantum field theory. The Vafa-Witten theorem \cite{VW} offers a strong constraint on the vacuum of vector-like gauge theories. Recently, It was shown that in a sufficiently strong magnetic field, the QCD vacuum can undergo a transition to a new phase when charged $\rho$ mesons condense\cite{EFTSC}. In this phase the vacuum behaves as an anisotropic inhomogeneous superconductor that supports superconductivity along the axis of the magnetic field and generates $\rho$ vortices. As these initial results are either from an effective theory for $\rho$ mesons\cite{EFTrho} or from a phenomenological NJL model \cite{NJLSC}, an investigation of the underlying fundamental QCD is needed. For QCD, the results seem to be controversial: paper \cite{QCDnegative} claims that in terms of the Vafa-Witten theorem, charged vector mesons cannot condense out in a magnetic field, whereas a later paper \cite{QCDpositive} states that this is not the case. Indeed, both papers agree that there is no Goldstone boson which originally is taken as a typical signature of global $U(1)_{I_3}$ isospin symmetry breaking in QCD for strong magnetic fields. The work \cite{QCDnegative} treats it as indicating an absence of condensation, whereas the work \cite{QCDpositive} explains it as the result of the Higgs mechanism. That is, due to the lock of the $U(1)_{I_3}$ with electro-magnetic $U(1)_{\mathrm{em}}$ gauge symmetry, charged $\rho$-meson condensation will induce spontaneous $U(1)_{\mathrm{em}}$ breaking; the photon will then obtain mass by 'eating' a Goldstone boson as its longitudinal component. Considering that the absence of a Goldstone boson can have different interpretations, the direct computation of a nonzero condensate becomes ever more important. Paper \cite{QCDnegative} explicitly performs this computation and shows that the condensate is zero, whereas paper \cite{QCDpositive} avoids directly discussing this issue. Nevertheless, nonzero charged $\rho$ meson condensation is preferred in effective theories and phenomenological model computations, and more importantly, intuition based on the important formula for the $\rho$ meson mass in an external magnetic field
\begin{eqnarray}
m^2_{\rho^\pm}(B_{\mathrm{ext}})=m^2_{\rho^\pm}-eB_{\mathrm{ext}}
 \end{eqnarray}
tells us that when the external magnetic field exceeds its critical value $m^2_{\rho^\pm}/e$, the effective $\rho$ mass $m_{\rho^\pm}(B_{\mathrm{ext}})$ becomes imaginary. This is a sign of an unstable vacuum similar to the famous situation for the electroweak Higgs potential; specifically, the curvature of the effective potential for a charged $\rho$ meson field near the origin (symmetric vacuum) is to be changed from concave to convex if the effective $\rho$ mass becomes imaginary. As long as the effective potential is bounded from below, convex behavior of the effective potential near origin demands that there should then exist at least two new non-symmetric vacuums characterized by nonzero condensates. One might guess that, in \cite{QCDnegative}, a no-condensate result is due to the fact that there are some unexpected infinities not counted in the computation that multiply the obtained zero creating a finite nonzero condensate. Unfortunately, up to now, this kind of infinity has not been found. One is then lead to question the reliability of the computation. The purpose of this paper is to show that this is really the case, i.e., the isospin violation condensation in QCD with $\theta=0$ might not be zero. There exists a loop-hole in the direct computation of the condensate either in the original Vafa-Witten's work or in the paper \cite{QCDnegative}.

The Vafa-Witten theorem shows that the vector-like global symmetries cannot be spontaneously broken in vector-like theories with zero theta angle. The result for continuous global symmetries is supported by two facts: direct computation shows a zero condensate for the order parameter, and a combination of various inequalities indicate no Goldstone boson. If both of these two facts are true, we call that the strong version of Vafa-Witten theorem holds.
 As discussed above, the absence of a Goldstone boson cannot be treated as a criterion for the spontaneous isospin symmetry breaking in QCD with strong external magnetic fields. In this special case, the validity of the strong version of Vafa-Witten theorem relies solely on the direct computation of the condensate in QCD and in the rest of the paper, we shall mainly focus on this issue.

In 2-flavor QCD with an external magnetic field, the charged $\rho$ meson condensate in terms of quark field $\psi$ is\footnote{Here, we perform calculations in Minkowski space, the rotation to Euclidian space being unnecessary for the present discussions.}
\begin{eqnarray}
\langle 0|\overline{\psi}(x)\tau_\pm\gamma_\pm\psi(x)|0\rangle_B\sim\overline{\mathrm{tr}\bigg[\gamma_\pm\tau_\pm
    ~\frac{1}{i\slashed{D}-m}\bigg](x,x)}\;,\label{condensate}
\end{eqnarray}
where $\tau$  and $\gamma$ are the Pauli and gamma matrices defined in the isospin and spinor spaces, respectively,  $\tau_\pm=(\tau_1\pm\tau_2)/2$ and $\gamma_{\pm}=(\gamma_1\pm i\gamma_2)/\sqrt{2}$. Subscript '$B$' denotes the background external magnetic field, 'tr' signifies the trace for isospin and spinor indices, $D_\mu=\partial_\mu-igT^aA^a_\mu-iqA^{\mathrm{em}}$ is the QCD covariant derivative for the gluon fields $A^a_\mu$ in the presence of an external $U(1)_{\mathrm{em}}$ field $A^{\mathrm{em}}_\mu$ with $g$ the strong coupling constant, and
$q=e(\tau_3+1/3)/2$ the electric charge for quarks. We take the u and d quarks with the same current mass $m$. The average in (\ref{condensate}) is computed for the gluon field with the standard QCD path integral measure; $\sim$ is to note that we have ignored the multiplication of some irrelevant finite nonzero constant. Because term $\frac{1}{i\slashed{D}-m}$ only has $1$ and $\tau_3$ components and no $\tau_\pm$ components in the isospin space, then $\mathrm{tr}\big[\gamma_\pm\tau_\pm~\frac{1}{i\slashed{D}-m}\big](x,x)=0$ due to fact that $\mathrm{tr}_{I_3}(\tau_\pm)=\mathrm{tr}_{I_3}(\tau_\pm\tau_3)=0$, where
$\mathrm{tr}_{I_3}$ is the trace of the isospin component. This situation is similar to isospin violation discussed in the original Vafa-Witten paper, where the condensate is
\begin{eqnarray}
\langle 0|\overline{\psi}(x)\tau_3\psi(x)|0\rangle\sim\overline{\mathrm{tr}\bigg[\tau_3
    ~\frac{1}{i\slashed{\tilde{D}}-m}\bigg](x,x)}\;,\label{condensate-1}
\end{eqnarray}
where $\tilde{D}_\mu=\partial_\mu-igT^aA^a_\mu$ is the pure QCD covariant derivative corresponding to the gluon fields. In (\ref{condensate-1}), the term $\frac{1}{i\slashed{\tilde{D}}-m}$ has no $\tau_3$ component in isospin space, hence $\mathrm{tr}\big[\tau_3~\frac{1}{i\slashed{\tilde{D}}-m}\big](x,x)=0$ due to the fact that $\mathrm{tr}_{I_3}(\tau_3)=0$. To avoid trivial zeroes, \cite{QCDnegative} adds to the Lagrangian an explicit infinitesimal isospin breaking perturbation term $\lim_{\epsilon\rightarrow 0}\epsilon\overline{\psi}\Gamma\psi$ with
 $\Gamma$ depending on isospins, spinors and space-time coordinate. Vafa-Witten does a similar thing by separating $m_u\neq m_d$ infinitesimally. Both studies show that the result is stable under infinitesimal isospin-violated perturbations. Thus, from direct computation giving vanishing condensates, \cite{QCDnegative} is at the same level as the original Vafa-Witten discussion.

To examine whether it is possible to avoid the above no-condenste results for (\ref{condensate}) and (\ref{condensate-1}), notice that in the effective theory and the NJL model, the covariant derivative $D_\mu$ includes the $\rho$ field as $\rho_\mu^{(a)}\tau_a$. This $\rho$ field dependence of the covariant derivative will generate $\tau_\pm$ and $\tau_3$ components in the $\frac{1}{i\slashed{D}-m}$ and $\frac{1}{i\slashed{\tilde{D}}-m}$, thereby prohibiting the appearance of a final zero. If one further argues that those effective theories and phenomenological models are not QCD, then we can do a similar thing in QCD by applying the technique we developed in Ref.\cite{WQ2000} to integrate out quark and gluon fields and integrate in bilocal colorless meson fields $\Phi(x,x')$ and $\Pi(x,x')$ exactly, modifying (\ref{condensate}) to
\begin{eqnarray}
\overline{\mathrm{tr}\bigg[\gamma_\pm\tau_\pm
    ~\frac{1}{i\slashed{D}-m}\bigg](x,x)}&\sim&\int\mathcal{D}\Phi\mathcal{D}\Pi
~\mathrm{tr}\bigg[\gamma_\pm\tau_\pm~\frac{1}{i\slashed{\partial}+q\slashed{A}^{\mathrm{em}}-m-\Pi}\bigg](x,x)~
    \mathrm{exp}\bigg\{\mathrm{Trln}(i\slashed{\partial}+q\slashed{A}^{\mathrm{em}}-m-\Pi)\nonumber\\
&&\hspace*{-1cm}+~i\int d^4x_1d^4x_1'N_c\Phi^{\sigma\rho}(x_1,x_1')\Pi^{\sigma\rho}(x_1,x_1')
+N_c{\displaystyle\sum_{n=2}}\int d^4x_1\cdots d^4x_nd^4x_1'\cdots d^4x_n'~\frac{(-i)^n(N_cg^2)^{n-1}}{n!}\nonumber\\
&&\times~\overline{G}^{\sigma_1\cdots\sigma_n}_{\rho_1\cdots\rho_n}
(x_1,x_1',\cdots,x_n,x_n')\Phi^{\sigma_1\rho_1}(x_1,x_1')\cdots\Phi^{\sigma_n\rho_n}(x_n,x_n')\bigg\}\;,\label{QCDEFF}
\end{eqnarray}
where $\sigma$ and $\rho$ are isospin and spinor index sets, Tr is the trace over color, isospin, spinor, and space-time indices, and $\overline{G}^{\sigma_1\cdots\sigma_n}_{\rho_1\cdots\rho_n}(x_1,x_1',\cdots,x_n,x_n')$ is the n-point gluon Green's function. For (\ref{condensate-1}), the corresponding result is just to replace $\gamma_\pm\tau_\pm$ in (\ref{QCDEFF}) by $\tau_3$ and to ignore the $q\slashed{A}^{\mathrm{em}}$ term. We see due to the appearance of isospin field $\Pi$ in $\frac{1}{i\slashed{\partial}+q\slashed{A}^{\mathrm{em}}-m-\Pi}$ that the original argument for a vanishing result now does not hold.

Suppose (\ref{QCDEFF}) does yield a nonzero result; one could still question whether (\ref{condensate}) and (\ref{condensate-1}) give zero results. Why does this no-condensate result, once the path integral is changed to some other form, become nonzero? To answer these questions, notice that once spontaneous symmetry breaking occurs, the order parameter should be a multi-valued quantity. However, (\ref{condensate}) and (\ref{condensate-1}) themselves  only give a single valued zero result for the order parameters $\langle 0|\overline{\psi}(x)\tau_\pm\gamma_\pm\psi(x)|0\rangle_B$ and $\langle 0|\overline{\psi}(x)\tau_3\psi(x)|0\rangle$; {\it the expected multi-value result does not show up}, while (\ref{QCDEFF}) does implies the existence of multi-value result. Here  the multivaluedness is associated with possible multiple local minima of the effective energy functional. These suggest that (\ref{condensate}) and (\ref{condensate-1}) are just results pertaining to the symmetric vacuum.
Without the knowledge of other possible vacuums, it is dangerous to jump to conclusion only with the information on the symmetric vacuum. To search for  multi-valued results, one needs to find some alternative expression which can include in the non-symmetric vacuum effect, like that given by (\ref{QCDEFF}). The standard way to deal with this situation is to introduce an effective action to search for different vacuums and assess which is physical according to its vacuum energy. In more detail, suppose $\phi(x)$ is an order parameter such as $\langle 0|\overline{\psi}(x)\tau_\pm\gamma_\pm\psi(x)|0\rangle_B$ or $\langle 0|\overline{\psi}(x)\tau_3\psi(x)|0\rangle$; introduce an external source term $J(x)\phi(x)$ into the Lagrangian of the path integral and construct the connected generating functional $W[J]$ on $J$ and classical field $\phi_c(x)=\frac{\delta W[J]}{\delta J(x)}$. The condensate is then the value of $\phi_c(x)$ when we switch off the external source $J=0$. This $\langle 0|\phi(x)|0\rangle=\phi_c\big|_{J=0}$ simply corresponds to the present direct computation of the condensate given in (\ref{condensate}) and (\ref{condensate-1}).

In practice, usually we do not know whether $W[J]$ is a single-valued functional on $J$ or not. We usually start computations by assuming a single-valued functional, hence $\phi_c(x)$ should be single valued. To retain multi-valued condensate solutions, instead of directly calculating $\phi_c$, one introduces an effective action $\Gamma[\phi_c]=W[J]-\int d^4x \phi_c(x)J(x)$, and expresses the effective action $\Gamma$ in terms of the classical field $\phi_c(x)$, not the original external source $J$. It is then easy to show $\frac{\delta\Gamma[\phi_c]}{\delta\phi_c(x)}=-J(x)$, which implies that $\phi_c(x)$ is an extremum of $\Gamma$. If $J=0$, one only needs to search for a real minimum of the effective potential (multiplied by the space-time volume and a minus sign, it is just the translational invariant part of the effective action) to obtain the physical condensation. For spontaneous symmetry breaking, with the exception of the conventional symmetric solution $\phi_c(x)\big|_{J=0}=0$, there usually exist other nonzero solutions and these correspond to lower vacuum energies. In this analysis to evaluate physical condensate, the key is that the external source should not vanish before obtaining the final physical condensate.
For (\ref{condensate}), the corresponding generating functional is
\begin{eqnarray}
e^{iW[J_\pm]}&=&\int\mathcal{D}A^a_\mu\mathcal{D}\overline{\psi}\mathcal{D}\psi~
\mathrm{exp}\bigg\{i\int d^4x~\bigg[-\frac{1}{4}G_{\mu\nu}^aG^{a\mu\nu}
+\overline{\psi}(i\slashed{D}\!-\!m\!+\!\gamma_\pm\tau_\pm J_\pm)\psi\bigg]\bigg\}\label{GenDef}\\
&=&\int\mathcal{D}A^a_\mu~
\mathrm{exp}\bigg\{\mathrm{Trln}[i\slashed{D}\!-\!m\!+\!\gamma_\pm\tau_\pm J_\pm]+i\int d^4x~[-\frac{1}{4}G_{\mu\nu}^aG^{a\mu\nu}]\bigg\}\;,\nonumber
\end{eqnarray}
where $G_{\mu\nu}^a$ is the gluon field strength and in the second equality we have integrated out the quark fields. The classical field $\rho_\pm(x)$ is defined as
\begin{eqnarray}
\rho_\pm(x)=\frac{\delta W[J_\pm]}{\delta J_\pm(x)}=
\frac{\int\!\mathcal{D}A^a_\mu\mathrm{tr}\bigg[\gamma_\pm\tau_\pm\frac{-i}{i\slashed{D}-m+\gamma_\pm\tau_\pm J_\pm}\bigg](x,x)\mathrm{exp}\bigg\{\mathrm{Trln}[i\slashed{D}\!-\!m\!+\!\gamma_\pm\tau_\pm J_\pm]+i\int\! d^4x[-\frac{1}{4}G_{\mu\nu}^aG^{a\mu\nu}]\bigg\}}{\int\mathcal{D}A^a_\mu~
\mathrm{exp}\bigg\{\mathrm{Trln}[i\slashed{D}\!-\!m\!+\!\gamma_\pm\tau_\pm J_\pm]+i\int d^4x~[-\frac{1}{4}G_{\mu\nu}^aG^{a\mu\nu}]\bigg\}}\;.\label{QCDEFTcondensate}
\end{eqnarray}
Note here the external-source-dependent isospin violation term $\gamma_\pm\tau_{\pm}J_\pm$ appearing in the denominator of the integrand in addition to the covariant derivative $\slashed{D}$. This makes the result nonzero as long as we do not let the external source $J_\pm$ vanish. The effective action is
\begin{eqnarray}
\Gamma[\rho_\pm]=W[J_\pm]-\int d^4x~\rho_\pm(x)J_\pm(x)\;,
\end{eqnarray}
which satisfies $\frac{\delta\Gamma[\rho_\pm]}{\delta\rho_\pm(x)}=-J_\pm(x)$. Here we emphasize that once we take $J_\pm=0$ at the very beginning in (\ref{GenDef}) and (\ref{QCDEFTcondensate}), as has been done in (\ref{condensate}) and (\ref{condensate-1}), we will just get the symmetric solution $\rho_\pm(x)\big|_{J_\pm=0}=0$. The nonzero solution can be obtained only when we come to calculate $\Gamma[\rho_\pm]$, which demands that one must not let the external source $J_\pm$ vanish in the process of computation. Instead, the external source must be replaced with its function on the classical field $J_\pm=J_\pm[\rho_\pm]$. Is this constraint on vanishing external source late so important? What is the effect if we replace it with its function on $\rho_\pm$? Note that to prove a theorem, we need to investigate all possibilities, while to show the theorem is not true,  just a counter example is enough. Here we exhibit an explicit example that zero condensate from directly calculation of generating functional, once considering its effective action, does lead physical nonzero condensations. This just contradicts with the logic of Vafa-Witten theorem that zero condensate from directly calculation of generating functional will lead physical zero condensation. To avoid QCD complexities and see clearly the core of the problem, we take this counter example as simple as possible, which will describe what happens and show strong version of the Vafa-Witten theorem may not to hold.

We consider a 0-dimension field model with generating functional $W[J]$ given as
\begin{eqnarray}
W[J]=12\lambda r^{\frac{4}{3}}\cos(\Theta+\frac{4}{3}\pi)[\cos(\Theta+\frac{4}{3}\pi)+\cos(3\Theta)]\;,\label{Wdef}
\end{eqnarray}
where $\lambda$ and $r$ are two real positive numbers of the model, quantity $\Theta$ relates these and the external source $J$ through
\begin{eqnarray}
3\Theta=\mathrm{arc~cos}\left(\frac{J}{8r\lambda}\right)\;.\label{thetaDef}
\end{eqnarray}
The classical field $\phi$ is defined as
\begin{eqnarray}
\phi=\frac{\delta W[J]}{\delta J}=2r^{\frac{1}{3}}\cos(\Theta+\frac{4}{3}\pi)\;,\label{CfieldDef}
\end{eqnarray}
where we have used the expression for $W[J]$ given in (\ref{Wdef}) and relation (\ref{thetaDef}) to obtain an explicit expression for the classical field $\phi$ in terms of $\Theta$. Notice that switching off the external source $J=0$ corresponds to
\begin{eqnarray}
\Theta\bigg|_{J=0}=\frac{\pi}{6}\;,\label{SymVac}
\end{eqnarray}
which allows the classical field to vanish, $\phi\big|_{J=0}=0$. This is a single-valued vanishing result, just as in (\ref{condensate}) or (\ref{condensate-1}). To generate nonzero $\phi\big|_{J=0}$, we first solve (\ref{CfieldDef}) and express the external source $J$ in terms of the classical field $\phi$ to obtain
\begin{eqnarray}
J=4\lambda(\phi^3-3r^{\frac{2}{3}}\phi)\;.\label{EQphi}
\end{eqnarray}
Noting that as our model is 0-dimensional, i.e., there is no space-time integration and the functional differential reverts to an ordinary derivative, we introduce the effective action as
\begin{eqnarray}
\Gamma=W-J\phi=6\lambda r^{\frac{2}{3}}\phi^2-\lambda\phi^4\;,\label{EFTphi4}
\end{eqnarray}
where, with the help of (\ref{EQphi}), we can perform some algebraic computations to express $\Gamma$ in terms of the classical field $\phi$. The result $\Gamma$ given by (\ref{EFTphi4}) is simply the 0-dimensional $\lambda\phi^4$ model with the wrong-sign mass-squared, $12\lambda r^{\frac{2}{3}}$. The vacuum is determined by $\frac{\delta\Gamma}{\delta\phi}=0$ and it leads to
 \begin{eqnarray}
 [\phi^2-3r^{\frac{2}{3}}]\phi=0\;.
 \end{eqnarray}
$\phi=0$ is the solution for the symmetrical vacuum, which is already predicted from (\ref{CfieldDef}) and (\ref{SymVac}) originally by directly computing the generating functional (\ref{Wdef}). The other two nonzero solutions $\phi=\pm \sqrt{3}r^{\frac{1}{3}}$ are new ones which cannot be obtained from (\ref{CfieldDef}) and (\ref{SymVac}). One can easily check that these two solutions have equal lower vacuum energies $V=-\Gamma|_{J=0}$ in comparison with the zero solution, hence correspond to physical condensates. This toy model thus produces nonzero condensates from a single-valued generating functional for the symmetrical vacuum, as long as we use the effective action formalism. The reason we call (\ref{Wdef}) and (\ref{thetaDef}) the generating functional on the symmetrical vacuum is that there exist two other generating functionals on non-symmetric vacuums by changing (\ref{thetaDef}) to
\begin{eqnarray}
3\Theta=\left\{\begin{array}{l}
2\pi-\mathrm{arc~cos}\left(\frac{J}{8r\lambda}\right)\\~\\
4\pi+\mathrm{arc~cos}\left(\frac{J}{8r\lambda}\right)
\end{array}\right.\;,\label{thetaDef-1}
\end{eqnarray}
which leads exactly to the same classical field expression (\ref{CfieldDef}) and the effective action (\ref{EFTphi4}). When we switch off the external source, $J=0$, instead of (\ref{SymVac}), these two generating functionals on the non-symmetric vacuums give
\begin{eqnarray}
\Theta\big|_{J=0}=\left\{\begin{array}{l}
\frac{\pi}{2}\\~\\
\frac{3\pi}{2}
\end{array}\right.\;,
\end{eqnarray}
which combined with (\ref{CfieldDef}) just give the nonzero solutions $\phi=\pm\sqrt{3}r^{\frac{1}{3}}$. Indeed, the three expressions from (\ref{thetaDef}) and (\ref{thetaDef-1}) just cover the three different solutions of equation (\ref{EQphi}), if we want solutions expressing $\phi$ in terms of the external source $J$. $\phi$ as function of $J$ is a three-valued function. In other words, it has three branches: the one given by (\ref{thetaDef}) corresponds to the symmetric branch, the other two given by (\ref{thetaDef-1}) to the non-symmetric branches. We refer to these results as those on symmetric and non-symmetric vacuums, respectively. Our original generating functional (\ref{Wdef}) just chooses the symmetric branch. This toy model indicates that due to the choice of a generating functional on the symmetrical vacuum, (\ref{condensate}) and (\ref{condensate-1}) produces a no-condensate result does not imply that the physical result must vanish.

One may criticize the simplicity of this toy model that first it is in 0-dimension space and then there is no motion for the particle;  second there is no path integral and then may be no quantum effect. To overcome these shortcomings, we generalize above 0-dimensional $\lambda\phi^4$ model to D-dimensional $O(N)$ $\lambda\phi^4$ model by considering following generating functional
\begin{eqnarray}
e^{i\tilde{W}_D[J]}=\int[{\mathcal D}\phi_i]~\exp\bigg\{i\int d^Dx~\bigg[\phi_i(x) [-\frac{1}{2}\partial_x^2+6\tilde{\lambda}\tilde{r}^{\frac{2}{3}}]\phi_i(x)-\frac{\tilde{\lambda}}{N}[\phi_i(x)\phi_i(x)]^2
+J_i(x)\phi_i(x)\bigg]\bigg\}\;,
\end{eqnarray}
where $\tilde{\lambda}$ and $\tilde{r}$ are two real positive numbers of the model. $\phi_i(x)$ has $N$ components, $J_i(x)$ is corresponding external source, $i=1,2,\ldots,N$. We discuss the large N limit of the model. Under this limit, there is an overall factor $N$ appears in the exponential of the integrand of the path integral due to summation of different components of $\phi(x)$ fields, which will suppress all loop contributions \cite{Jackiw}
\begin{eqnarray}
\tilde{W}_D[J]\stackrel{N\rightarrow\infty}{====\Rightarrow}
\int d^Dx~\bigg[\phi_{i,c}(x) [-\frac{1}{2}\partial_x^2+6\tilde{\lambda}\tilde{r}^{\frac{2}{3}}]\phi_{i,c}(x)
-\frac{\tilde{\lambda}}{N}[\phi_{i,c}(x)\phi_{i,c}(x)]^2
+J_i(x)\phi_{i,c}(x)\bigg]
\end{eqnarray}
where $\phi_{i,c}(x)$ is the solution of equation
\begin{eqnarray}
\frac{\delta}{\delta \phi_{i,c}(x)}
\int d^Dy~\bigg[\phi_{i,c}(y) [-\frac{1}{2}\partial_y^2+6\tilde{\lambda}\tilde{r}^{\frac{2}{3}}]\phi_{i,c}(y)
-\frac{\tilde{\lambda}}{N}[\phi_{i,c}(y)\phi_{i,c}(y)]^2
+J_i(y)\phi_{i,c}(y)\bigg]=0\;,
\end{eqnarray}
which leads,
\begin{eqnarray}
J_i(x)=\bigg[\frac{4\tilde{\lambda}}{N}\phi_{j,c}(x)\phi_{j,c}(x)+(\partial_x^2-12\tilde{\lambda}\tilde{r}^{\frac{2}{3}})\bigg]\phi_{i,c}(x)\label{JiDef}\;.
\end{eqnarray}
Above equation is difficult to solve, we consider a translational invariant situation that external source $J_i$ and
$\phi_{i,c}$ are all independent of space-time coordinates. Then above equation is reduced to
\begin{eqnarray}
\tilde{J}=\bigg[\frac{4\tilde{\lambda}}{N}\tilde{\phi}^2-12\tilde{\lambda}\tilde{r}^{\frac{2}{3}}\bigg]\tilde{\phi}\;,\label{mod}
\end{eqnarray}
where $\tilde{J}=\sqrt{J_iJ_i}$ and $\tilde{\phi}=\sqrt{\phi_{i,c}\phi_{i,c}}$ are modulus of $J_i$ and $\phi_{i,c}$ in $O(N)$ space respectively. Above equation is the same as Eq.(\ref{EQphi}) if we identify $\tilde{J}$, $\tilde{\phi}$, $\tilde{\lambda}/N$, $N\tilde{r}^{2/3}$ with $J$, $\phi$, $\lambda$, $r^{2/3}$, respectively. This equation has three solutions if we want to express the classical field $\tilde{\phi}$ in terms of the external source $\tilde{J}$. Among them, one solution cannot satisfy positivity requirement for $\tilde{J}$ and $\tilde{\phi}$, the left two solutions are
\begin{eqnarray}
\tilde{\phi}=2N^{\frac{1}{2}}r^{\frac{1}{3}}\cos(\tilde{\Theta}+\frac{4}{3}\pi)\;,
\end{eqnarray}
and
\begin{eqnarray}
3\tilde{\Theta}
=\left\{\begin{array}{lll}
\mathrm{arc~cos}\left(\frac{\tilde{J}}{8N^{\frac{1}{2}}\tilde{r}\tilde{\lambda}}\right)&&\mbox{symmetric branch}\\~\\
2\pi-\mathrm{arc~cos}\left(\frac{\tilde{J}}{8N^{\frac{1}{2}}\tilde{r}\tilde{\lambda}}\right)&~~~~&\mbox{non-symmetric branch}
\end{array}\right.\;,
\end{eqnarray}
The translational invariant generating functional of large N limit D-dimensional $O(N)$ $\lambda\phi^4$ model becomes
\begin{eqnarray}
&&\hspace{-1cm}\tilde{W}_D[J]\stackrel{N\rightarrow\infty;~J_i(x)=J_i}{=========\Rightarrow}
12N\tilde{\lambda}\tilde{r}^{\frac{4}{3}}\cos(\tilde{\Theta}+\frac{4}{3}\pi)
[\cos(\tilde{\Theta}+\frac{4}{3}\pi)+\cos(3\tilde{\Theta})]\int d^Dx\;.
\end{eqnarray}
We see that path integral now generates two different branch generating functionals, one is for symmetric vacuum,  another is for non-symmetric vacuum. If one chooses the symmetric one as \cite{QCDnegative} and \cite{VW} at the very beginning of the computation for condensate, it will lead vanishing result. The difference with \cite{QCDnegative} and \cite{VW} is that now instead of single value generating functional, path integral gives multi-value result. The reason present direct computation for this D-dimensional $O(N)$ $\lambda\phi^4$ model resulting in  multi-value result is that path integral involves the degree of freedom of order parameter $\phi_i$, just like (\ref{QCDEFF}). While in \cite{QCDnegative} and \cite{VW}, due to choice of fundamental quark as dynamical variable and the lack of explicitly extracting out the degree of freedom of order parameter, the generating functional unfortunately choose the symmetric branch.

In more general situations, if we assume the effective action $\Gamma[\phi]$ be a single-valued function, then for spontaneous symmetry breaking, solving equation $\frac{\delta\Gamma[\phi]}{\delta \phi(x)}=-J(x)$ to express $\phi(x)$ in terms of $J$ will usually lead to a multi-valued function, or a function which has several branches, because spontaneous symmetry breaking demands multi-vacuum solutions. Among these branches, one is the symmetric branch corresponding to the generating functional on the symmetric vacuum. The other branches lead to generating functionals on non-symmetric vacuums. The no-condensate results obtained in \cite{QCDnegative} and Vafa-Witten's paper \cite{VW} is due to choosing the generating functional on the symmetric vacuum from the very beginning. The proofs in
 \cite{VW} and \cite{QCDnegative} do not deny the other nonsymmetric vacuum (or other branch of the generating functional), while our expression (\ref{QCDEFF}) explicitly show the possibility of existence of other nonsymmetric vacuum. Then just the discussion on the symmetric vacuum may not represent the correct  result of the whole system. To avoid that misleading result, the safest way is to calculate the effective action, or changing over to the generating functional on the non-symmetric vacuum as done in an effective theory or the NJL model or (\ref{QCDEFTcondensate}), where the path integral is already improved involving degree of freedom relate to order parameter as our D-dimensional $O(N)$ $\lambda\phi^4$ model. Taking just an infinitesimal perturbation, as Vafa-Witten and Ref.\cite{QCDnegative} had done, is not enough to cross-over from the symmetric vacuum to the non-symmetric vacuum.

With this discussion on direct computation of condensate, we find that the no-condensate result in the Vafa-Witten theorem and \cite{QCDnegative} may not be true due to its prejudiced choice of a generating functional on symmetric vacuum and lack the information for other possible nonsymmetric vacuum. Ignoring this possible defect, the Vafa-Witten theorem has to reduce to its weaker form, i.e., it only proves that in vector-like gauge theories with $\theta=0$, there is no Goldstone boson. The charged $\rho$ meson condensate in QCD for strong magnetic fields happen to have no Goldstone boson due to existence of Higgs mechanism.

The explanation of generating functional on symmetric vacuum may also help to understand the dispute over the parity violation problem in QCD \cite{Parity}. Once we accept the fact that the generating functional Vafa-Witten used in their proof is the one associated with the symmetric vacuum, then the absence of a Goldstone boson obtained from this generating functional is also questionable. Considering that up to now, unlike the vanishing condensate case, there is no exceptions to the result of no Goldstone boson in various effective field theories and phenomenological models, this partial result may still be true. In fact,  one can check that original Vafa-Witten's discussion on two-point correlation function of isospin current for QCD in a strong magnetic field is still true in presence non-vanishing isospin source introduced as in Eq.(\ref{GenDef}) and the exponential fall-off behavior for the correlation function remains, thus insuring the correctness of the weak version of the Vafa-Witten theorem.

It should be emphasized that our discussions are not only valid for the Vafa-Witten theorem, but also apply to all kinds of direct computations of various physical quantities and all generating functional-based results. That is, once the system has spontaneous symmetry breaking or multi-vacuums, one must carefully check the results obtained from the direct computation of path integrals to see that these are not performed just on the non-physical vacuum.

\section*{ACKNOWLEDGMENTS}
We thank Professor Xiaoyuan Li for recommending the present topic to us. This work is supported by the National Science Foundation of China (NSFC) under Grants No.11075085, Specialized Research Fund Grants No.20110002110010 for the Doctoral Program of High Education of China, and Tsinghua University Initiative Scientific Research Program.

 
\end{document}